\documentclass[12pt]{iopart}

\usepackage{iopams}
\usepackage{color}

\usepackage[utf8]{inputenc}       
\usepackage[english]{babel} 
\usepackage{dsfont}
\usepackage[perpage,symbol*]{footmisc}
\usepackage[final]{graphicx}                             
\begin{document}

\title[A quaternionic unification of electromagnetism and hydrodynamics]{A quaternionic unification of electromagnetism and hydrodynamics}

\author{Arbab I. Arbab}

\address{Department of Physics, Faculty of Science, University of Khartoum, P.O. Box 321, Khartoum
11115, Sudan}
\ead{aiarbab@uofk.edu}

\begin{abstract}
We have derived energy conservation equations from the quaternionic Newton's law that is compatible with Lorentz transformation. This Newton's law yields directly the Euler equation and other equations governing the fluid motion. With this formalism, the pressure contributes positively to the dynamics of the system in the same way mass does. Hydrodynamic equations are derived from Maxwell's equations by adopting an electromagnetohydrodynamics (EMH) analogy. In this analogy the hydroelectric field is related to the local acceleration of the fluid and the Lorentz gauge is related to the incompressible fluid condition. An analogous Lorentz gauge in hydrodynamics is proposed. We have shown that the vorticity of the fluid is developed whenever the particle local acceleration of the fluid deviates from the velocity direction. We have also shown that Lorentz force in electromagnetism corresponds to Euler force in fluids. Moreover, we have obtained Gauss's, Faraday's and Ampere's -like laws in Hydrodynamics.
\end{abstract}
\pacs{41.20.Jb; 47.10.-g; 47.10.A-}
\maketitle
\section{Introduction}
We have recently formulated Maxwell equations using quaternion (Arbab and Satti, 2009 \& Arbab, 2009). In this formalism, we have shown that Maxwell equations have  not included the magnetic field produced by the charged particle. Consequently, we have also shown that this magnetic field is given by an equation equivalent to Biot-Savart law. Moreover, we have shown that the magnetic field created by the charged particle is always perpendicular to the particle direction of motion. We have then  introduced the quaternionic continuity equation and compared it with the ordinary continuity equation. Consequently, we have found that Maxwell equations predict that the electromagnetic fields propagate in vacuum or a charged medium with the same speed of light in vacuum if this quaternioic continuity equation is valid.

In the present paper we explore the idea of formulating the fluid  equations using the quaternion description. Owing to the previous analysis done in electromagnetism, we would like to further establish an analogy between quaternionic hydrodynamic and quaternionic Maxwell's equations.
Accordingly, it turns out that the quaternionic Newton's equation gives directly the non-relativistic limit of the energy momentum conservation equation.  These equations are the Euler equation, the continuity equation, the energy momentum conservation equation. Thus, an analogy between hydrodynamics and electrodynamics is shown to exist. This analogy guarantees that one can derive the appropriate equations of the former system from the latter or vice versa. This analogy is quite impressing since it allows us to visualize the flow of the electromagnetic field as resembling the flow of a fluid.  A quaternionic Newton's second law is developed and a generalized Newton's law of motion is then obtained.  The application of the  generalized continuity equations (GCE) to the generalized Newton's law yields  the non-relativistic equations governing the motion of fluids. It is interesting to notice that  generalized continuity equations are Lorentz invariant. The fluid vorticity is found to arise whenever the local acceleration of the fluid particles deviates from the velocity direction. The physical properties and the equations governing the hydrodynamics are derived. We remark here that these equations can be put in in analogous forms to Gauss's, Faraday's and Ampere's laws. Moreover, we have found that the diffusion equation is compatible with the GCEs. We therefore, emphasize that the GCEs should append any model dealing with fluid motion. According to this analogy and since the electrodynamics is written in terms of the electric and magnetic fields, $\vec{E}_e$ and $\vec{B}$ respectively, we should write the hydrodynamics equations in terms of the hydroelectric field $\vec{E}_h$  and vorticity $\vec{\omega}$ and will then obtained analogous equations to Maxwell equations. This is legitimate because of the analogy that exists between the two paradigms permits, viz.,   $\vec{E}_h\Leftrightarrow \vec{E}_e$  and $\vec{\omega}\Leftrightarrow \vec{B}$, $\vec{v}\Leftrightarrow \vec{A}$ and $\chi=\frac{v^2}{2}\Leftrightarrow \varphi$, where $\vec{E}_h=-\frac{\partial \vec{v}}{\partial t}$, $\vec{\omega}=\vec{\nabla}\times \vec{v}$.

\section{The Continuity equation}
We have recently  explored the application of quaternions  to Maxwell's equations (Arbab and Satti, 2009). We have found that the quaternionic Maxwell's equation reduces to the ordinary Maxwell's equations but predicts the existence of a scalar wave competing with the electromagnetic field traveling at the speed of light in vacuum.  Quaternions are found to have interesting properties. The multiplication rule for the two quaternions, $\widetilde{A}=(a_0\,, \vec{A})$ and $\widetilde{B}=(b_0\,, \vec{B})$ is given by (Tait, 1904)
\begin{equation}
\widetilde{A}\widetilde{B}=\left(a_0b_0-\vec{A}\cdot\vec{B}\,\,, a_0\vec{B}+\vec{A}b_0+\vec{A}\times\vec{B}\right).
\end{equation}
Therefore, the ordinary continuity equation is transformed into a quaternionic continuity equation as

\begin{equation}
\widetilde{\nabla}\widetilde{J}=\left[-\,\left(\vec{\nabla}\cdot
\vec{J}+\frac{\partial \rho}{\partial t}\right)\, ,\,
\frac{i}{c}\,\left(\frac{\partial \vec{J}}{\partial
t}+\vec{\nabla}\rho\, c^2\right)+\vec{\nabla}\times
\vec{J}\,\right]=0,
\end{equation}
where
\begin{equation}
\widetilde{\nabla}=\left(\frac{i}{c}\frac{\partial}{\partial
t}\, , \vec{\nabla}\right)\,,\qquad \widetilde{J}=\left(i\rho c\, ,
\vec{J}\right)\,.
\end{equation}
The scalar and vector parts of Eq.(2) imply that
\begin{equation}
\vec{\nabla}\cdot \vec{J}+\frac{\partial \rho}{\partial
t}=0\,,
\end{equation}
\begin{equation}
\vec{\nabla}\rho+\frac{1}{c^2}\frac{\partial
\vec{J}}{\partial t}=0\,,
\end{equation}
and
\begin{equation}
\vec{\nabla}\times
\vec{J}=0\,.
\end{equation}
We call Eqs.(4)-(6) the generalized continuity equations (GCEs).
In a covariant form, Eqs.(5)-(6) read
\begin{equation}
\partial_\mu J^\mu=0\,,\qquad \partial_\mu J_\nu-\partial_\nu J_\mu=0\,.
\end{equation}
Hence the GCEs are Lorentz invariant. We remark that the GCEs are applicable to any flow whether created by charged particles or neutral ones.
\section{ Newton's second law of motion:}
The motion of the mass ($m$) is governed by the Newton's second law. The quaternionic Newton force reads (Arbab and Satti, 2009)
\begin{equation}
\widetilde{F}=-m\widetilde{V}(\widetilde{\nabla}
\widetilde{V})\,,\,\,\,\,
\end{equation}
where
\begin{equation}
\widetilde{F}=\left(\frac{i}{c}P\,,\vec{F}\right) , \qquad \widetilde{V}=\left(ic\,\,,\, \vec{v}\right).
\end{equation}
The vector part of Eq.(8) yields the two equations
\begin{equation}
\vec{F}=m\left(\frac{\partial \vec{v}}{\partial
t}+\vec{\nabla}\left(\frac{v^2}{2}\right)-\vec{v}\times\left(\vec{\nabla}\times\vec{v}\right)\right),
\end{equation}
and
\begin{equation}
\vec{\nabla}\times\vec{v}=-
\frac{\vec{v}}{c^2}\times\frac{\partial
\vec{v}}{\partial t}\,.
\end{equation}
The scalar part of Eq.(8) yields the two equations
\begin{equation}
P =mc^2\left(\vec{\nabla}\cdot\vec{v}+\frac{\vec{v}}{c^2}\cdot\frac{\partial
\vec{v}}{\partial t}\right)\,,
\end{equation}
and
 \begin{equation}
\vec{v}\cdot\left(\vec{\nabla}\times\vec{v}\right)=0\,.
\end{equation}
For a continuous medium (fluid) containing a volume $V$, one can write Eq.(10) as
\begin{equation}
\rho\left(\frac{\partial \vec{v}}{\partial
t}+\vec{\nabla}(\frac{v^2}{2})-\vec{v}\times(\vec{\nabla}\times\vec{v})\right)=\vec{f}\,\,,
\end{equation}
where $m=\rho \, V$ and $\vec{f}=\frac{\vec{F}}{V}$\,.
Using the vector identity $\frac{1}{2}\vec{\nabla} (\vec{v}\cdot \vec{v})=\vec{v}\times(\vec{\nabla}\times \vec{v})+(\vec{v}\cdot\vec{\nabla})\,\vec{v}$, Eq.(14) becomes
\begin{equation}
\rho\left(\frac{\partial \vec{v}}{\partial
t}+(\vec{v}\cdot\vec{\nabla})\,\vec{v}\right)=\vec{f}\,\,.
\end{equation}
This is the familiar Euler equation describing the motion of a fluid.
For a fluid moving under pressure ($P_r$), one can write the pressure force density as
\begin{equation}
\vec{f}_P=-\vec{\nabla} P_r\,,
\end{equation}
so that Eq.(15) becomes
\begin{equation}
\rho\left(\frac{\partial \vec{v}}{\partial
t}+(\vec{v}\cdot\vec{\nabla})\,\vec{v}\right)=-\vec{\nabla} P_r\,\,.
\end{equation}
Using Eq.(5), Eq.(12) can be written as
\begin{equation}
\frac{\partial u}{\partial t}+\vec{\nabla}\cdot\vec{S}=\vec{f}\cdot \vec{v}\,,\qquad \vec{S}=(\rho c^2)\,\vec{v}, \qquad u=\rho\,v^2\,,\qquad  P=\vec{f}\cdot \vec{v}\,.
\end{equation}
Eq.(18) is an energy conservation equation, where $\vec{S}$ is the energy flux and $u$ is the energy density of the moving fluid.
With pressure term only, Eq.(18) yields
\begin{equation}
\frac{\partial u}{\partial t}+\vec{\nabla}\cdot(\rho\,c^2+P_r)\,\vec{v}=P_r\vec{\nabla}\cdot\vec{v}\,.
\end{equation}
The source term on the right hand side in the above equation is related to the work needed to expand the fluid. It is shown by Lima {\it et al.} (1997) that such a term has to be added to the usual equation of fluid dynamics to account for the work related to the local expansion of the fluid. It is thus remarkable we derive the fundamental hydrodynamics equations from just two simple  quaternionic equations, the continuity and Newton's equation.
For  incompressible fluids $\vec{\nabla}\cdot\vec{v}=0$ so that Eq.(19) becomes
\begin{equation}
\frac{\partial
u}{\partial t}+\vec{\nabla}\cdot(\rho\,c^2+P_r)\,\vec{v}=0\,,
\end{equation}
which states that the pressure contributes  positively to the energy flow. This means the total energy flow of the moving fluid is
\begin{equation}
\vec{S}_{\rm total}=(\rho \,c^2+P_r)\, \vec{v}\,,
 \end{equation}
and the total momentum density of the flow is given by
\begin{equation}
\mathfrak{\vec{p}}=(\rho+\frac{P_r}{c^2})\,\vec{v}\,,
 \end{equation}
which must be conserved. This is analogous to the general theory of relativity where the pressure and mass are sources of gravitation. Equation (20) states also that there is no loss or gain of energy for incompressible fluids. However, when viscous terms considered loss of energy into friction  will arise. In standard cosmology the general trend of introducing the bulk viscosity ($\eta$)\footnote{The viscous pressure can be obtained from the viscous force, $\vec{F}=-A\,\eta\, \frac{d\vec{v}}{dr}\Rightarrow P_v=\frac{F}{A}=-\eta \, \vec{\nabla}\cdot\vec{v}$}  is  by replacing the pressure term $P_r$ by the effective  pressure (Weinberg, 1972)
\begin{equation}
 P_{\rm eff.} =P_r-\eta\vec{\nabla}\cdot\vec{v}\,.
\end{equation}
Substituting this in Eq.(17) and defining the vorticity of the fluid by $\vec{\omega}=\vec{\nabla}\times \vec{v}$, we get
\begin{center}
$$
\hspace{3cm}\rho\left(\frac{\partial \vec{v}}{\partial
t}+(\vec{v}\cdot\vec{\nabla})\,\vec{v}\right)=-\vec{\nabla} P_r+\eta\nabla^2\vec{v}+\eta \vec{\nabla}\times\vec{\omega}\,\,,
  \hspace{2.0cm}\qquad (16a)
 $$
\end{center}
which reduces to the Navier-Stokes equation for irrotational flow ($\vec{\omega}=0$).
Equation (20) can be put in a covariant form as
\begin{equation}
\partial_\mu T^{\mu\nu}=0\,, \qquad T_{\mu\nu}=(\rho+\frac{P_r}{ c^2})\,v_\mu v_\nu-P_rg_{\mu\nu}\,.
\end{equation}
where $T_{\mu\nu}$ is the energy momentum tensor of a perfect fluid, $v_\mu$ is its velocity and $g_{\mu\nu}$ is the metric tensor with signature (+++-). It is interesting to remark that we pass from quaternion Newton's law to relativity without any offsetting. This is unlike the case of ordinary Newton's law where relativistic effects can't be included directly.

Using the vector identity, $\vec{\nabla}\times(f\vec{A})=f(\vec{\nabla}\times \vec{A})-\vec{A}\times(\vec{\nabla}f)$, with  $\vec{J}=\rho\, \vec{v}$, Eq.(6) can be written as
\begin{equation}
\vec{\nabla}\times\vec{J}=\vec{\nabla}\times (\rho\, \vec{v})=\rho\left(\vec{\nabla}\times\vec{v}\right)-\vec{v}\times(\vec{\nabla}\rho)=0\,,
\end{equation}
which upon using Eq.(4) transforms into
\begin{equation}
\vec{\nabla}\times\vec{v}=-
\frac{\vec{v}}{c^2}\times\frac{\partial
\vec{v}}{\partial t}\,.
\end{equation}
Thus, Eq.(10) derived from Newton's second law  is equivalent to one of the continuity equations, viz., Eq(5).
Taking the dot product of Eq.(5) with a constant velocity $\vec{v}$, we get
$$
\vec{v}\cdot\vec{\nabla}\rho+\frac{1}{c^2}\vec{v}\cdot\frac{\partial
\vec{J}}{\partial t}=0\,,
$$
which yields
$$
\frac{d\rho}{d\,t}=\frac{\partial}{\partial t}(1-\frac{v^2}{c^2})\,\rho\,.
$$
According to Lorentz transformation, if the density in the rest frame is $\rho$, it will be $\rho'=(1-\frac{v^2}{c^2})\,\rho$ in the moving inertial frame. Thus, taking the total derivative, is equivalent to taking the partial derivative of the density in a moving inertial frame (Lawden, 1968).
In terms of the vorticity, the above equation becomes
\begin{equation}
\vec{\omega}=
\frac{\vec{v}}{c^2}\times\left(-\frac{\partial
\vec{v}}{\partial t}\right)\,.
\end{equation}
The vorticity is related to the angular velocity of the fluid ($\vec{\Omega}$) by the relation $\vec{\omega}=2\,\vec{\Omega}$.
 This clearly shows that the fluid motion is governed by the continuity equation as well as the Newton's equation. \\
In a recent paper, we have shown that the magnetic field produced by a moving charged particle due to an external electric field is given by (Arbab and Satti, 2009)
\begin{equation}
\vec{B}=\frac{\vec{v}}{c^2}\times\vec{E}\,.
\end{equation}

We remark here that there seems to be a resemblance between the vorticity of flow  and the magnetic field produced by the charged particle. This analogy is evident from the fact that $\vec{B}=\vec{\nabla}\times \vec{A}$ and $\vec{\omega}=\vec{\nabla}\times \vec{v}$. Moreover,  Eq.(13) shows that the fluid helicity, $h_f=\vec{v}\cdot\vec{\omega}=0$, and Eq.(28) shows that the magnetic helicity, $h_m=\vec{v}\cdot\vec{B}=0$ (Kikuchi, 2007). Using vector identities, it is obvious from Eq.(11) that $\vec{\nabla}\cdot \vec{\omega}=0$. This equation resembles the equation $\vec{\nabla}\cdot \vec{B}=0$. The former equation implies that vortex lines must form closed loops or be terminated at a boundary, and that the strength of a vortex line remains constant. Thus, a charged particle creates a magnetic field associated  with the particle in the same way as the fluid creates a vortex that moves with the particle. According to De Broglie hypothesis, a wave nature is associated with all moving microparticles. It, thus, seems that vorticity and magnetic fields travel like a wave.

With the same token, Eqs.(27) and (28) suggest that the hydroelectric field is given by
\begin{equation}
\vec{E}_h=-\frac{\partial \vec{v}}{\partial t}\qquad\Rightarrow\qquad \vec{\omega}=\frac{\vec{v}}{c^2}\times \vec{E}_h\,.
\end{equation}
This field is generated due to the fluid (mass) motion. Moreover, the Coulomb  gauge ($\vec{\nabla}\cdot \vec{A}=0)$ in electrodynamics is equivalent to the incompressibility of the fluid ($\vec{\nabla}\cdot \vec{v}=0)$, however, the Lorentz gauge is equivalent to
\begin{equation}
\vec{\nabla}\cdot \vec{v}+\frac{1}{c^2}\frac{\partial (v^2/2)}{\partial t}=0
\end{equation}
 in hydrodynamics\footnote{This gauge when multiplied by a constant density $\rho$ yields the momentum conservation equation.}. This latter equation shows that the incompressibility is maintained for a steady flow only. When the fluid expands, this merit is lost and the Lorentz gauge applies once again.
  However, from our above definition of hydroelectric field, we must impose the condition that $\vec{\nabla}\chi=0$, i.e., $\chi$ is spatially independent but can depend on time, i.e., $\chi=\chi(t)$. The unit of  $\chi$ is $J/kg$, or $m^2/s^2$. This defines the energy required to move one kilogram of fluid. If we had defined the hydroelectric field in Eq.(29) as $\vec{E}_h=-\frac{\partial \vec{v}}{\partial t}-\vec{\nabla}\frac{v^2}{2}$ (where  $\chi=\frac{v^2}{2}$), the Euler force in  Eq.(10) would become
\begin{equation}
\vec{F}_h=-m\left(\vec{E}_h+\vec{v}\times\vec{\omega}\right)\,,
\end{equation}
which is equivalent to Lorentz force in electromagnetism
  \begin{equation}
\vec{F}_{em}=q\left(\vec{E}_e+\vec{v}\times \vec{B}\right)\,,
\end{equation}
This is a very interesting analogy, since for electron  $q<0$  so that $-m<0$, $\vec{B}\Leftrightarrow\vec{\omega}$   and $\vec{E}_e\Leftrightarrow\vec{E}_h$. The complete analogy is tabulated below. We would like to call this symmetry an electromagnetohydrodynamics (EMH) analogy. The following table shows the analogy between electromagnetic and hydrodynamics. As the electric field of an electron points opposite to the force direction, the hydroelectric field ($\vec{E}_h$) points opposite to the direction of flow motion.
 It is an amazing analogy. Employing this analogy, we would like to derive the hydrodynamics laws from the electrodynamics corresponding ones.
We call this symmetry an electromagnetohydrodynamics (EMH) transitivity. The following table shows the analogy between electromagnetic and hydrodynamics. As the electric field of an electron points opposite to the force direction, the hydroelectric field ($\vec{E}_h$) points opposite to the direction of flow motion.
\\
\\
\begin{tabular}{|l|l|l|l|l|l|}
  \hline

  Theory & Circulation & Gauge fields & Gauge condition & Helicity  & Electric field\\

  \hline
   \emph{Electrodynamics} & $\vec{B}=\vec{\nabla}\times \vec{A}$ & $\varphi,\,\, \vec{A}$  & $\vec{\nabla}\cdot \vec{A}+\frac{1}{c^2}\frac{\partial \varphi}{\partial t}=0$& $h_e=\vec{v}\cdot \vec{B}$  & $\vec{E}_e=-\frac{\partial \vec{A}}{\partial t}-\vec{\nabla}\varphi$
   \\

    \emph{Hydrodynamics} & $\vec{\omega}=\vec{\nabla}\times \vec{v}$ & $\chi=\frac{v^2}{2},\,\, \vec{v}$   & $\vec{\nabla}\cdot \vec{v}+\frac{1}{c^2}\frac{\partial \chi}{\partial t}=0$ & $f_h=\vec{v}\cdot \vec{\omega}$  & $\vec{E}_h=-\frac{\partial \vec{v}}{\partial t}-\vec{\nabla}\chi$  \\
  \hline
\end{tabular}
\\
\\
\\
It is an amazing analogy. Employing this analogy, we would like to derive the hydrodynamics laws from the electrodynamics corresponding ones.
The Maxwell's equations are
\begin{equation}\label{2}
\vec{\nabla}\times \vec{B}=\mu_0 \vec{J}+\frac{1}{c^2}\frac{\partial \vec{E}_e}{\partial t},
\end{equation}
\begin{equation}\label{2}
\vec{\nabla}\times \vec{E}_e+\frac{\partial \vec{B}}{\partial t}=0,
\end{equation}
\begin{equation}\label{2}
\vec{\nabla}\cdot\vec{E}_e=\frac{\rho_e}{\varepsilon_0},
\end{equation}
and
\begin{equation}\label{2}
\vec{\nabla}\cdot \vec{B}=0\,,
\end{equation}
where $\rho_e$ is the charge density.
Employing the EMH analogy, taking the divergence of both sides of Eq.(33) and  employing Eq.(4) yields
\begin{equation}\label{2}
\vec{\nabla}\cdot \vec{E}_h=\frac{\rho_m}{\varepsilon_h}\,,
\end{equation}
 where   $\mu_h$ replaces $\mu_0$  and $c_s$  replaces $c$ for the  EMH, where $\mu_h=\frac{\kappa c_s^2}{\gamma^2}$, where $\gamma$  is the surface tension and $\kappa$  is the bulk modulus of the fluid under question. Notice that the unit of $\mu_0$  is $Hm^{-1}$, $\varepsilon_0$  is $Fm^{-1}$  and the constant $a_0$  is $kg m^{-1}$  , where $a_0=\frac{\gamma^2}{c_s^2\kappa}$ , which may define the resistance (fluidity)  of the fluid to flow.  Equation (37) is nothing but Gauss law in hydrodynamics, where $\vec{E}_e\rightarrow \vec{E}_h$  and the mass density ($\rho_m$ ) replaces the charge density ($\rho_e$). Besides, Eq.(37) implies the hydrodynamic permittivity, $\varepsilon_h=\frac{\gamma^2}{\kappa c_s^4}$. Hence,  $\mu_h\varepsilon_hc_s^2=1$ (in comparison with $\mu_0\varepsilon_0c^2=1$). Notice however that one can relate this constant to Newton's constant by the relation  $a_0=\frac{c_s^2}{4\pi G}$. In this case Eq.(37) becomes
  \begin{equation}\label{2}
\vec{\nabla}\cdot \vec{E}_h=4\pi G\rho_m\,,
\end{equation}
 which also suggests that $G=\frac{1}{4\pi \epsilon_h}$ , which is to be compared with Coulomb constant $k=\frac{1}{4\pi \varepsilon_0}$. Therefore, the constant  $a_0$ is a new fundamental constant. One can therefore define the gravitational permittivity, $\varepsilon_h=1.19\times 10^{9}\rm kgm^{-3}s^2$  and the gravitational permeability as $\mu_h=\frac{4\pi G}{c_s^2}=7.25\times 10^{-15}\rm kg m^{-1}$. In such a case the electric field defined in Eq.(38) will be gravitoelectric field. This coincides with the value proposed recently by Merches and Onuta by considering an ideal gravitomagnetic fluid (Merches and Onuta, 1998). The vorticity field intensity of a moving fluid is $\vec{H}_f=\frac{\vec{\omega}}{\mu_h}$.
For the Earth  $\omega=10^{-14}s^{-1}$ so that $H_f=1.0\, \rm kgm^{-1}s^{-2}$.
Equation (38) can be compared with the gravitational field ($\vec{g}$) equation due to a static mass distribution of the form, $\vec{\nabla}\cdot \vec{g}=-4\pi G\rho_m$, where  $\rho_m$ is a function of space variables only. In this formalism, we see that a spinning or orbiting object will generate a hydroelectromagnetic field, where the hydroelectric field is equal to the centripetal acceleration.
The time dependence of the hydroscalar  $\chi$  can be obtained from Eq.(38) using Eqs.(29) and (30) so that
 \begin{equation}\label{2}
\frac{1}{c_s^2}\frac{\partial^2\chi}{\partial t^2}=\frac{\rho_m}{\varepsilon_h}\,,\qquad \chi=\frac{v^2}{2}\,.
\end{equation}
 This equation will have direct consequences for cosmological applications where  varies with cosmic time. For an expanding universe, where $a$ is the scale factor of the universe, $\vec{\nabla}\cdot \vec{v}=3\left(\frac{\dot a}{a}\right)=3H$ and $H$  is the Hubble constant. Hence, upon using Eqs.(2) and (30), Eq.(39) yields the relation $\rho_m\propto t^{-2}$ which is a solution of Friedmann equation describing the cosmological expansion of a perfect fluid (the universe) (Weinberg, 1972). We remark here that Eq.(39) can be applicable to the motion of air inside the lung parenchyma.
The velocity vector field ( $\vec{v}$) can then be obtained by solving Eq.(38) and applying the solution in Eq.(29). The solution of Eq.(39) depends on how  $\rho_m$ varies with time.
Using the EMH analogy, Eq.(33) yields
\begin{equation}\label{2}
\vec{\nabla}\times\vec{\omega}=\mu_h\vec{J}+\frac{1}{c_s^2}\frac{\partial\vec{E}_h}{\partial t}\,.
\end{equation}
This represents Ampere's law in hydrodynamics. Such a result is obtained by (Sulaiman  and   Handoko) in their analogy between electomagnetsism and fluid mechanics [13]. Now take the time derivative of Eq.(27) and use Eq.(29) to get
$$\hspace{3cm}
\frac{\partial \vec{\omega}}{\partial t}=\frac{\vec{v}}{c_s^2}\times\left(\frac{\partial
\vec{E}_h}{\partial t}\right)\,.\hspace{8cm} (40a)
$$
Now use Eq.(40) to get
$$\hspace{3cm}
\frac{\partial \vec{\omega}}{\partial t}=\vec{v}\times(\vec{\nabla}\times\vec{\omega}-\mu_h\vec{J})\,,\hspace{4cm} \vec{J}=\rho\,\,\vec{v}\,.\hspace{2cm} (40b)
$$
Using the vector identity, $\vec{\nabla}(\vec{A}\cdot\vec{B})=\vec{A}\times(\vec{\nabla}\times\vec{B})+\vec{B}\times(\vec{\nabla}\times\vec{A})+(\vec{A}\cdot\vec{\nabla})\vec{B}
+(\vec{B}\cdot\vec{\nabla})\vec{A}$, and Eq.(13), we obtain  the equation of  vorticity for inviscid flow
$$\hspace{3cm}
\frac{\partial \vec{\omega}}{\partial t}+(\vec{v}\cdot\vec{\nabla})\,\vec{\omega}+(\vec{\omega}\cdot\vec{\nabla})\,\vec{v}=0\,.\hspace{6cm} (40c)
$$

It is interesting to note that taking the curl of Eq.(29) and using the fact that $\vec{\omega}= \vec{\nabla}\times\vec{v}$, one gets
\begin{equation}\label{2}
\vec{\nabla}\times\vec{E}_h+\frac{\partial\vec{\omega}}{\partial t}=0\,.
\end{equation}
This is the Faraday analogue of hydrodynamics, where $\vec{E}_h\Leftrightarrow \vec{E}_e$  and $\vec{\omega}_h\Leftrightarrow \vec{B}$. For steady flow, $\vec{E}_h=0$ and hence $\vec{\omega}=\rm const.$. This means that the vorticity is conserved.
Application of the EMH analogy to Eq.(36) yields
\begin{equation}\label{2}
\vec{\nabla}\cdot\vec{\omega}=0\,,
\end{equation}
which is true using the vector identity $\vec{\nabla}\cdot(\vec{\nabla}\times\vec{v})=\vec{\nabla}\cdot\vec{\omega}=0$.
Equations (38), (40), (41) and (42) represent the gravitoelectromagnetic field equations.
It is shown by Peng that Einstein field equations reduce in the limit of weak field to equations similar to these but with negative mass (Peng, 1983, 1990). However, our derivation here doesn't address this bizarre situation. This makes our model realistic and physically admissible.
Using Eq.(40), the fact that  $\vec{\omega}=\vec{\nabla}\times\vec{v}$ and the vector identity,  $\vec{\nabla}\times(\vec{\nabla}\times\vec{v})=\vec{\nabla}(\vec{\nabla}\cdot\vec{v})-\nabla^2\vec{v}$, one obtains
\begin{equation}\label{2}
\frac{1}{c_s^2}\frac{\partial^2\vec{v}}{\partial t^2}-\nabla^2\vec{v}=0\,.
\end{equation}
Equation (43) shows that the velocity vector field, $\vec{v}$, is a solution of a wave equation traveling at speed of sound. This has the same form as the equation governing the electric potential vector $\vec{A}$ in free space. Hence, instead of solving Euler equation, one can solve the above equation to get the velocity.  Differentiating of Eq.(43) partially with respect to time and using Eq.(29), one obtains
\begin{equation}\label{2}
\frac{1}{c_s^2}\frac{\partial^2\vec{E}_h}{\partial t^2}-\nabla^2\vec{E}_h=0\,.
\end{equation}
This equation shows that the hydroelectric field propagates with speed of sound in air, and having a source in the density gradient. Hence, the hydroelectric field will arise whenever there is any non-uniform spatial distribution of the density of the fluid. It is interesting that a static mass distribution can produce a hydroelectric field. It is not clear here whether this wave nature of the hydroelectric field is connected with the generation of gravitational wave. However, one may assume that any accelerating or non-uniform density distribution of the fluid (mass) will induce a hydroelectric field. The detection of such a field is to be studied elsewhere.

 Now, taking the cross product (from left) of both sides of Eq.(40) and using Eqs.(41), (42) and (6) the vector identity $\vec{\nabla}\times(\vec{\nabla}\times\vec{\omega})=\vec{\nabla}(\vec{\nabla}\cdot\vec{\omega})-\nabla^2\vec{\omega}$, one gets
\begin{equation}\label{2}
\frac{1}{c_s^2}\frac{\partial^2\vec{\omega}}{\partial t^2}-\nabla^2\vec{\omega}=0\,.
\end{equation}
This shows that the vorticity vector is a solution of a wave equation traveling at speed of sound in air.  It is evident that unlike the hydroelectric field the vorticity has no source. It is also similar to the evolution of the magnetic field $\vec{B}$. Equation (45) can also be obtained by taking the curl of Eq.(43). This equation has a spherically symmetric solution of the form
\begin{equation}\label{2}
\vec{\omega}(r,t)=\frac{\vec{C}(t-r/c_s)}{r}\,,
\end{equation}
where $\vec{C}(t-r/c_s)$  is an arbitrary vector of retarded time $t-r/c_s$ . The retardation $r/c_s$  is equal to the time needed for the vorticity wave to pass the distance from the source to a given point in space at a distance $r$. This exhibits the causal behavior associated with the wave disturbance. The argument of  $\vec{C}$ shows that an effect observed at the point  $r$ at time  $t$ is caused by the action of the source
a distant  $r$ away at an earlier or retarded time $t'=t-r/c_s$. The time $r/c_s$  is the time of propagation of the disturbance from the source to the point $r$.

According to Maxwell's equations the Poynting vector is given by
\begin{equation}\label{2}
\vec{S}_{em}=\frac{\vec{E}\times\vec{B}}{\mu_0}\,,
\end{equation}
so that in hydrodynamics,  Eq.(47) will become
\begin{equation}\label{2}
\vec{S}_{h}=\frac{\vec{E}_h\times\vec{\omega}}{\mu_h}\,,
\end{equation}
using Eq.(29), Eq.(48) yields
\begin{equation}\label{2}
\vec{S}_{h}=(\rho_m c_s^2)\vec{v}\,.
\end{equation}
This is the same as Eq.(18) with $P_r=0$. The vanishing of the pressure is due to the fact that $\vec{\nabla}\chi=0$.
Thus, it is remarkable that one can obtain all analogous formulae in gravitation by invoking the EMH analogy without deriving them.
\section{Concluding remarks}
We have studied in this paper the consequences of the quaternionic Newton's second law of motion. We have derived the energy conservation equation and Euler equation. These laws are compatible with the generalized continuity equations. The basic equations governing the fluid dynamics are also derived. We have then shown that the energy momentum equation is Lorentz invariant. Ordinary Newton's second law is known not to be compatible with Lorentz transformation. However, the generalized Newton's second laws are compatible with special relativity. Moreover, we have shown that the pressure contributes equally to the energy density of the moving fluid as the mass does. Besides, the pressure is the source of the hydroelectric scalar field. Moreover, we have found an intimate analogy between electrodynamics and hydrodynamics. With this analogy, we derived the magnetohydrodynamics equations from Maxwell's equations. We have obtained from Ampere's law, a Gauss's-like law applicable to gravitational (or hydrodynamic) system. Moreover, Faraday's-like law and Ampere's -like law in hydrodynamics are obtained. In this analogy Euler's force corresponds to Lorentz's force.
\section*{ACKNOWLEDGEMENTS}
I would like to thank F. Amin for stimulating discussion.
\section*{REFERENCES}
$[1]$ Arbab, A. I., and Satti, Z. A., "On the generalized Maxwell equations and their prediction of electroscalar wave", Progress in Physics, 2, 8 (2009).\\
$[2]$ Arbab, A. I.,   "On the new gauge transformations of Maxwell equations", Progress in Physics, 2, 14 (2009).\\
$[3]$ Tait, Peter Guthrie,  \emph{An elementary treatise on quaternions}, 2nd  ed., Cambridge  University Press (1873);
Kelland, P. and Tait, P. G., \emph{Introduction to Quaternions}, 3rd ed. London: Macmillan, (1904).\\
$[4]$ Lima, J.A.S., Zanchin, V, and Brandenberger, R., "On the Newtonian cosmology equations with pressure'', Mon. Not. R. Astron. Soc., 291, L1, (1997).\\
$[5]$ Kikuchi, H., {\it Progress In Electromagnetics Research Symposium}, Beijing, China, March 26-30, 996, (2007).\\
$[6]$ Davidson, P.A., An Introduction to Magnetohydrodynamics, Cambridge University Press, (2001).\\
$[7]$ Lawden, D.F., Tensor Calculus and Relativity, Methuen, London, (1968).\\
$[8]$ Weinberg, S., Introduction to Gravitation and Cosmology,  John Wiley and Sons, (1972).\\
$[9]$ Merches, I., and Onuta, T., "An analytical formulation in the theory of gravitomagnetic systems'', Analele Sttiintifice Ale Universitaii, Fizica Teoretica,  1997-1998.\\
$[10]$ Peng, H., "A new approach to studying local gravitomagnetic effects on a superconductor", Gen. Rel. Gravit., 22, 609 (1990).\\
$[11]$ Peng, H., "On calculation of magnetic-type gravitation and experiments", Gen. Rel. Gravit., 15, 725 (1983).\\
$[12]]$ Arbab, A. I., Gen. Rel. Gravit. 36, 2465 (2004).\\
$[13]$   Sulaiman, A., and   Handoko, L.T., http://arxiv.org/abs/physics/0508092v1.\\
\end{document}